\documentclass[
    ,final            % use final for the camera ready runs
  ]
  {aipproc}

\layoutstyle{6x9}

\newbox\grsign \setbox\grsign=\hbox{$>$}
\newdimen\grdimen \grdimen=\ht\grsign

\newbox\laxbox \newbox\gaxbox
\setbox\gaxbox=\hbox{\raise.5ex\hbox{$>$}\llap
        {\lower.5ex\hbox{$\sim$}}}\ht1=\grdimen\dp1=0pt
\setbox\laxbox=\hbox{\raise.5ex\hbox{$<$}\llap
        {\lower.5ex\hbox{$\sim$}}}\ht2=\grdimen\dp2=0pt

\newcommand{\ga}{\mathrel{\copy\gaxbox}}

\begin{document}

%\title{Radio lobes and X-ray hot spots of the extraordinary 
%microquasar in NGC\,7793}
\title{Radio lobes and X-ray hot spots of the 
extraordinary microquasar in NGC\,7793}

\classification{97.80.Jp}
\keywords      {X-ray binaries; Jets, outflows, and bipolar flows}

\author{Roberto Soria}{
  address={MSSL, University College London, Holmbury St Mary, Surrey RH5 6NT, UK}
}

\author{Manfred Pakull}{
  address={Observatoire Astronomique
de Strasbourg, 11 rue de l'Universit\'e, 67000 Strasbourg,
France}
}

\author{Jess Broderick}{
  address={School of Physics \& Astronomy,      
University of Southampton, Hampshire SO17 1BJ, UK}
%  ,altaddress={<author1 address>} % additional visiting address
}

\author{Stephane Corbel}{
  address={Universit\'e Paris 7 and Service d'Astrophysique, 
UMR AIM, CEA Saclay, F-91191, Gif sur Yvette, France}
}

\author{Christian Motch}{
  address={Observatoire Astronomique
de Strasbourg, 11 rue de l'Universit\'e, 67000 Strasbourg,
France}
}

\begin{abstract}
We have studied the newly-discovered microquasar in NGC\,7793 
in radio, optical and X-ray bands. This system 
comprises a large ($250 \times 120$ pc) line-emitting optical nebula, 
detected in H$\alpha$ and He{\footnotesize{II}} $\lambda 4686$. 
The optical nebula coincides with a synchrotron-emitting radio cocoon, 
with a radio luminosity about 4 times that of Cas A. 
The central black hole appears as a hard X-ray source with 
a point-like, blue optical counterpart. Two prominent radio lobes 
are located at the extremities of the cocoon. Just ahead of the radio 
lobes, we found two X-ray hot spots, which we interpret as a signature 
of the bow shock into the interstellar medium. The X-ray hot spots, 
radio lobes, X-ray core and major axis of the cocoon are well aligned, 
proving that the system is powered by a jet. From the X-ray data, we estimate 
a jet power $\sim$ a few $\times 10^{40}$ erg s$^{-1}$, active 
over a timescale $\approx 10^{5}$ yrs. This extraordinary system 
is a long-sought analog of the Galactic microquasar SS\,433.
\end{abstract}

\maketitle

%%%%%%%%%%%%%%%%%%%%%%%%%%%%%%%%%%%%%%%%%%%%
%% MAINMATTER
%%%%%%%%%%%%%%%%%%%%%%%%%%%%%%%%%%%%%%%%%%%%

\section{Radio galaxies and microquasars}

The basic physical model for radio lobes in FRII radio galaxies 
is based on a pair of relativistic, collimated jets emerging 
from the active black hole (BH). As the jet interacts with and is decelerated 
by the ambient (interstellar or intergalactic) medium, a reverse shock 
propagates inwards into the ejected plasma. After crossing the reverse 
shock, the jet material inflates a cocoon of hot gas, which is 
less dense but much overpressured with respect to the undisturbed 
medium. Thus, the cocoon expands supersonically, 
driving a forward shock (bow shock) into the ambient medium 
\citep{sch74,bla74,raw91,kai99}.   

The cocoon and lobes are the main sources of optically-thin 
(steep spectrum) synchrotron radio emission, while we expect 
optically-thick (flat-spectrum) radio emission from 
the jet near the core. A radio- and sometimes X-ray-luminous 
hot spot is usually found at the reverse shock, at the end 
of the jet. This is where 
most of the bulk kinetic energy of the jet is transferred 
to a non-thermal population of ultra-relativistic electrons, 
which cool via synchrotron and synchrotron self-Compton 
emission. 
%\citep{mei89,har04}. 
This scenario implies 
that the peaks of the radio and X-ray emission are spatially 
coincident. Optically-thin thermal plasma X-ray emission 
may come instead from the hot, shocked ambient gas in the thin 
layer between the cocoon and the bow shock, when its expansion 
is highly supersonic. In this case, the peak of the thermal 
X-ray emission will appear just in front of the radio lobe, 
as is the case for example in the nearest radio galaxy, 
Cen A \citep{kra07}.

There is a scale invariance between the jet emission processes
in microquasars (powered by stellar-mass BHs)
and in AGN/quasars (powered by supermassive BHs). 
%BH mass,
%core radio luminosity (a proxy for jet power) and X-ray luminosity
%identify a ``fundamental plane'' of BH accretion \citep{mer03,fen04}.
There is also at least one important difference: microquasars are
mostly located in a relatively low-pressure medium as compared
to the medium around AGN, when scaling of the jet
thrust is taken into account \citep{hei02}. 
As a consequence, we expect to see fewer, dimmer cocoons and radio lobes 
in microquasars than in the most powerful AGN and quasars;  
%(e.g., Fanaroff-Riley type II radio galaxies).
but the linear size of those microquasar cocoons and jets 
can be up to 1000 times larger than in radio galaxies, 
scaled to their respective BH masses.

So far, our knowledge of the interaction of microquasar jets
with the interstellar medium has largely relied on the
Galactic microquasar SS\,433 and its surrounding synchrotron-emitting
nebula W50 (size $\sim 100 \times 50$ pc).
A mildly relativistic ($v_J = 0.27c$), precessing jet acts
as a sprinkler that inflates ``ear-like'' structures,
protruding from the more spherical W50 nebula. Most of the jet
power ($\sim 10^{39}$--$10^{40}$ erg s$^{-1}$)
is dissipated in those lobes \citep{beg80}.
Faint evidence of the interaction of relativistic jets
with the interstellar medium have been found in a few
other, less powerful Galactic microquasars (e.g., Cyg X-1: \citep{gal05};
Cyg X-3: \citep{mar05}; GRS\,1915$+$105: \citep{kai04}).
On a larger scale, huge (size $\ga 100$ pc) ionized nebulae 
have been found around a significant fraction of ultraluminous
X-ray sources in nearby galaxies \citep{pak06,pak08,gri08}. 
Such nebulae emit optical lines
typical of shock-ionized gas, and in a few cases,
synchrotron radio emission \citep{mil05,sor06}. The derived ages
($\ga$ a few $10^5$ yrs) and energy content ($\sim 10^{52}$--$10^{53}$ erg)
are too large for ordinary supernova remnants, and suggest jet/wind inflation
with a mechanical power $\sim 10^{39}$--$10^{40}$ erg s$^{-1}$,
comparable with the X-ray luminosities \citep{pak06}.
However, no direct X-ray or radio evidence of a collimated jet 
had been found, until recently.

\begin{figure}
  \includegraphics[height=.27\textheight]{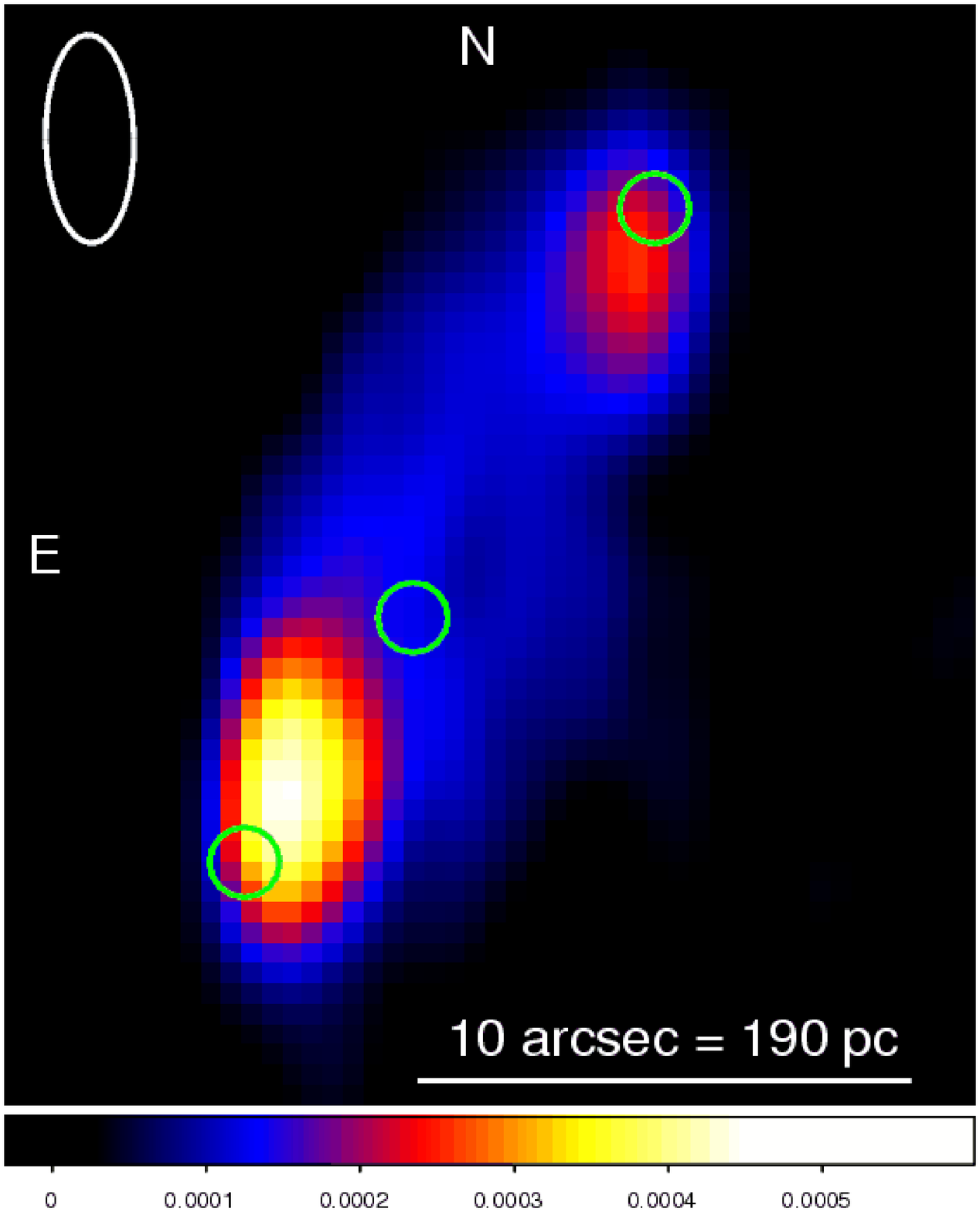}
  \includegraphics[height=.27\textheight]{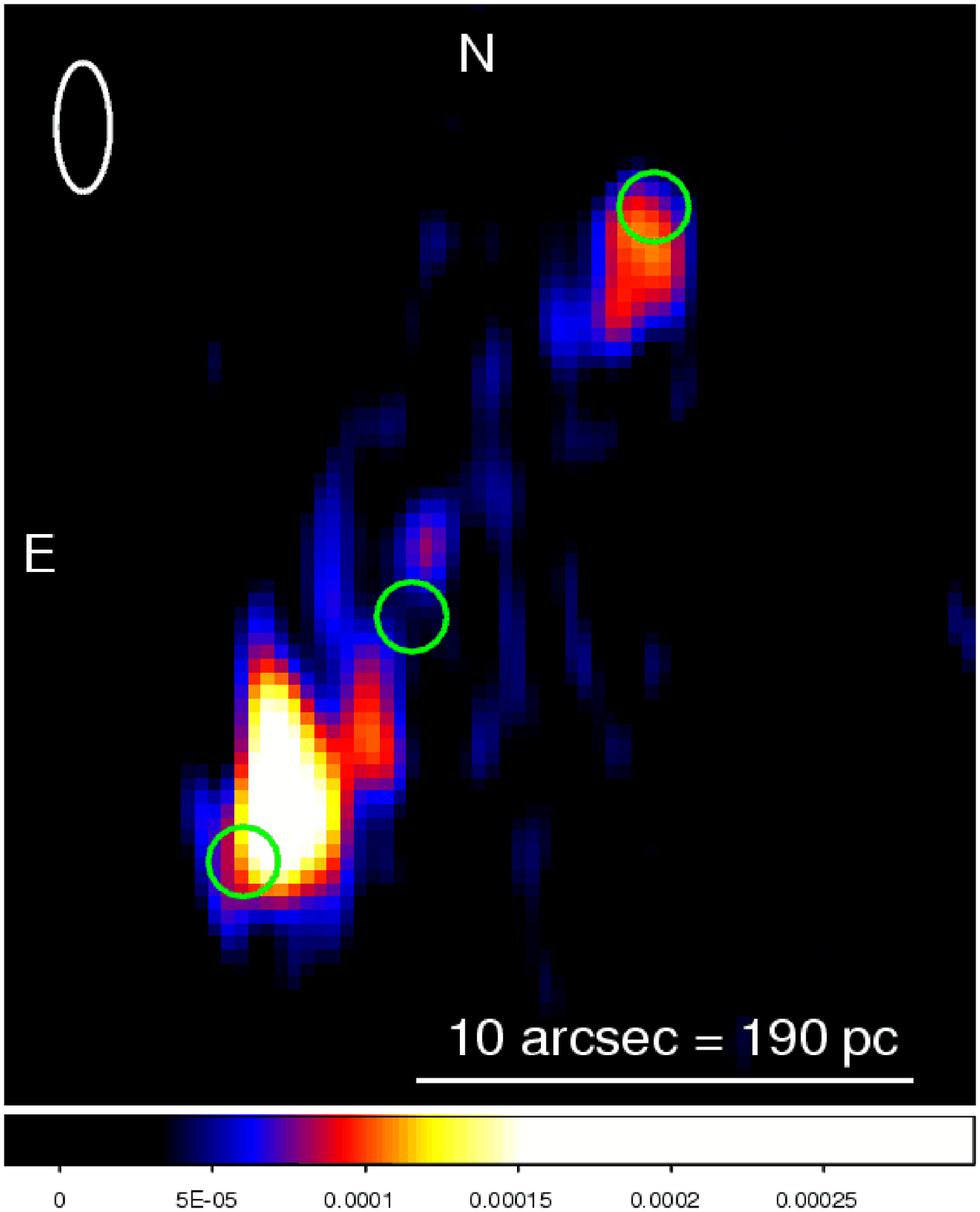}
  \includegraphics[height=.27\textheight]{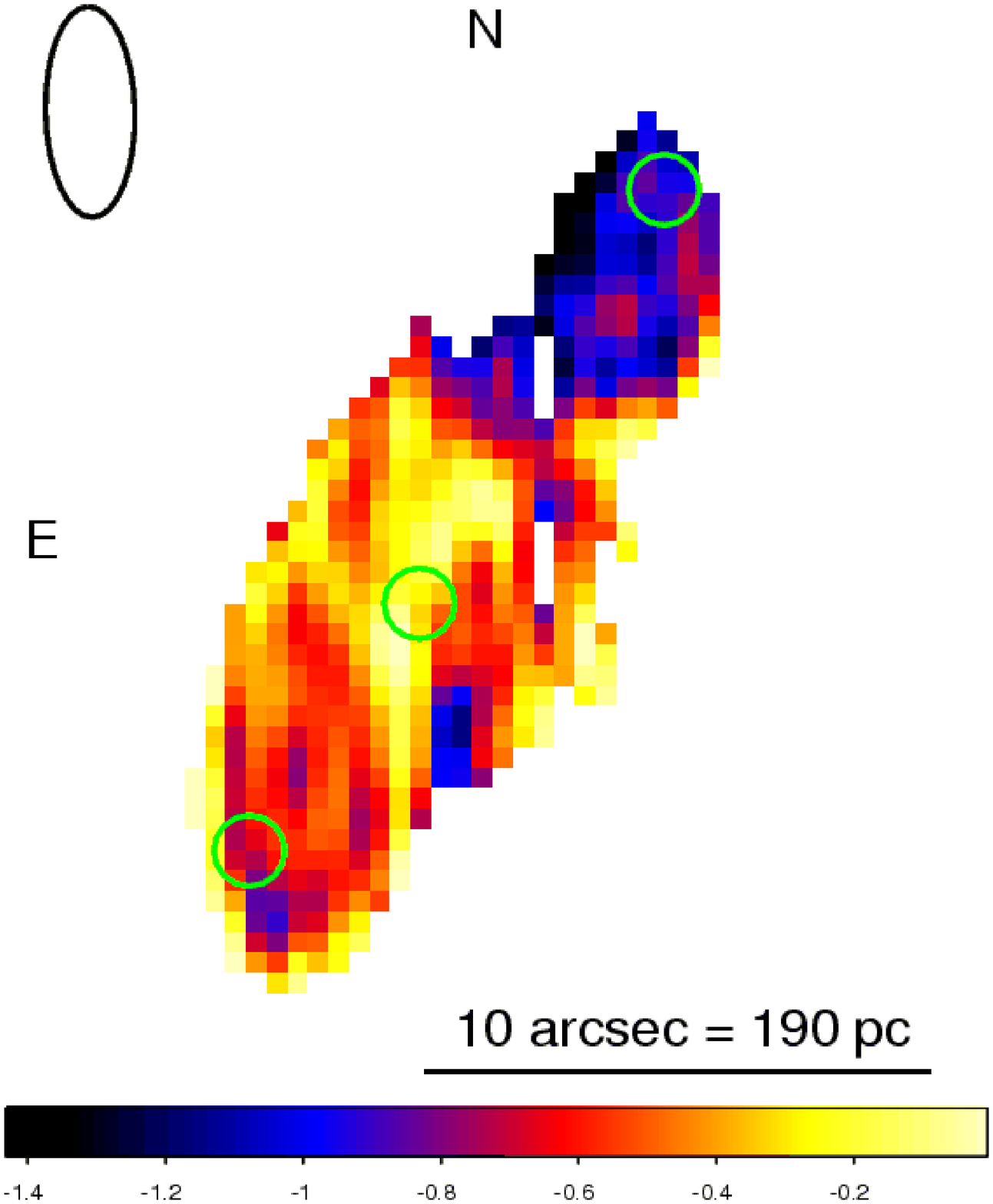}
  \caption{Left panel: ATCA map at 5.5 GHz, with flux scale 
   in Jy beam$^{-1}$. The overplotted green circles mark 
the position of the X-ray core and hot spots, from {\it Chandra}.
Middle panel: ATCA map at 9 GHz. Right panel: map of the 
radio spectral index, inferred from the ratio of the 5.5 GHz 
and 9 GHz maps, rescaled to the same beam size. Note the two flat-spectrum 
spurts at the base of the jet.}
\end{figure}

\section{The newly-discovered SS433-like microquasar}

An extraordinary radio/optical/X-ray microquasar
was recently discovered \citep{pak08} in the Sculptor 
galaxy NGC\,7793 ($d \approx 3.9$ Mpc). It has:
a large ($\approx 250 \times 120$ pc) shock-ionized nebula,
emitting in the radio bands, H$\alpha$ and He{\footnotesize{II}} 
$\lambda 4686$; prominent 
FRII-like radio lobes, more luminous than the rest of the cocoon; 
an X-ray core, presumably the location of the active BH; 
a blue optical counterpart to the X-ray core; 
a pair of X-ray hot spots, aligned with the radio lobes 
and the X-ray core, typical signature of energy transport 
via a collimated jet interacting with the interstellar medium. 
%The radio nebula has an integrated flux
%of $\approx 1.5$ mJy at 6 cm (more luminous than Cas A). 
The system resembles the famous Galactic microquasar SS\,433, 
but on an even grander scale. It was previously classified 
as the unusual supernova remnant S26 in \citep{bla97} and \citep{pan02}.
%We dubbed this system Cygnet A because it is a baby version 
%of the prototypical radio galaxy Cygnus A.

\begin{figure}
  \includegraphics[height=.25\textheight]{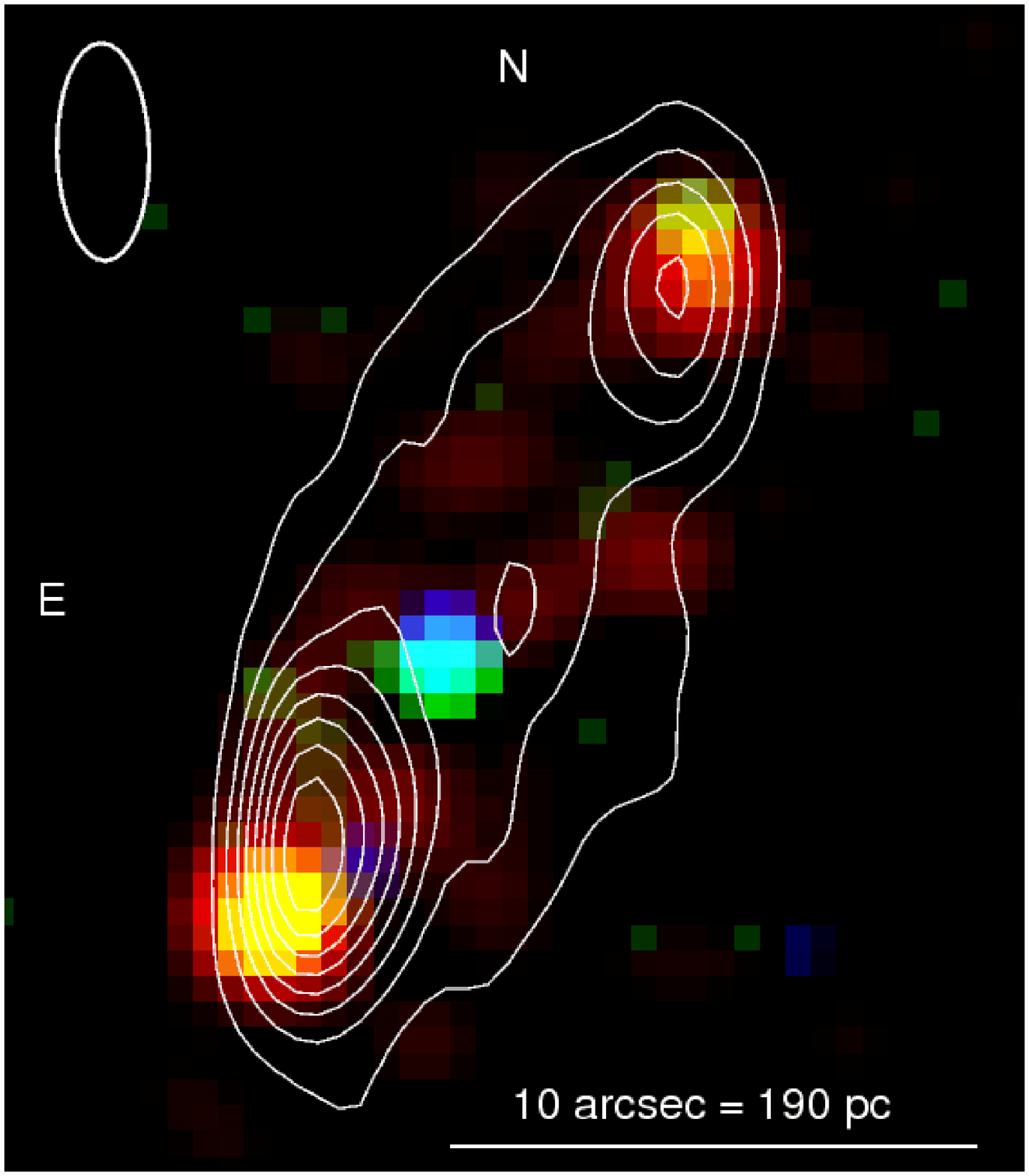}
  \includegraphics[height=.25\textheight]{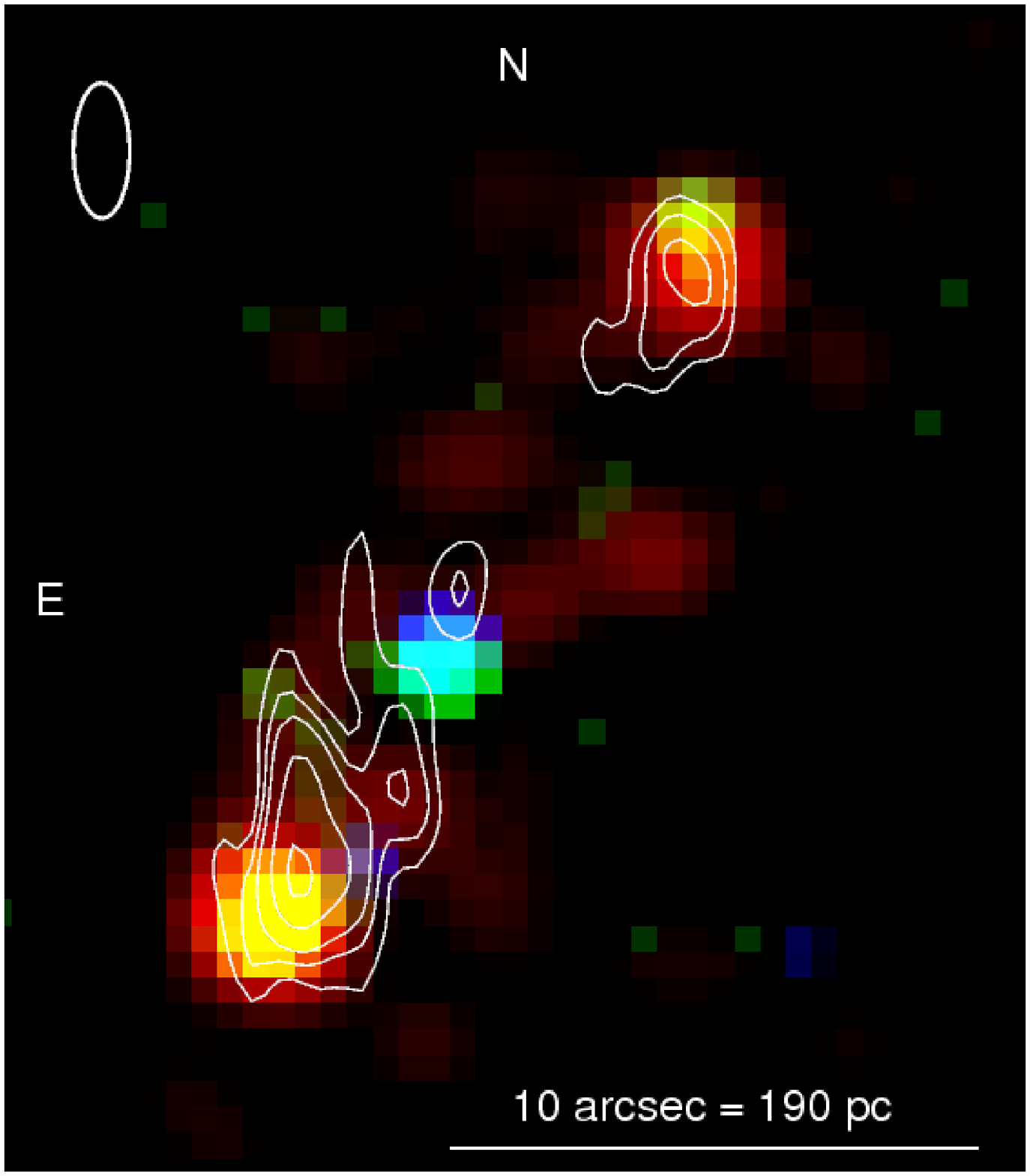}
  \includegraphics[height=.25\textheight]{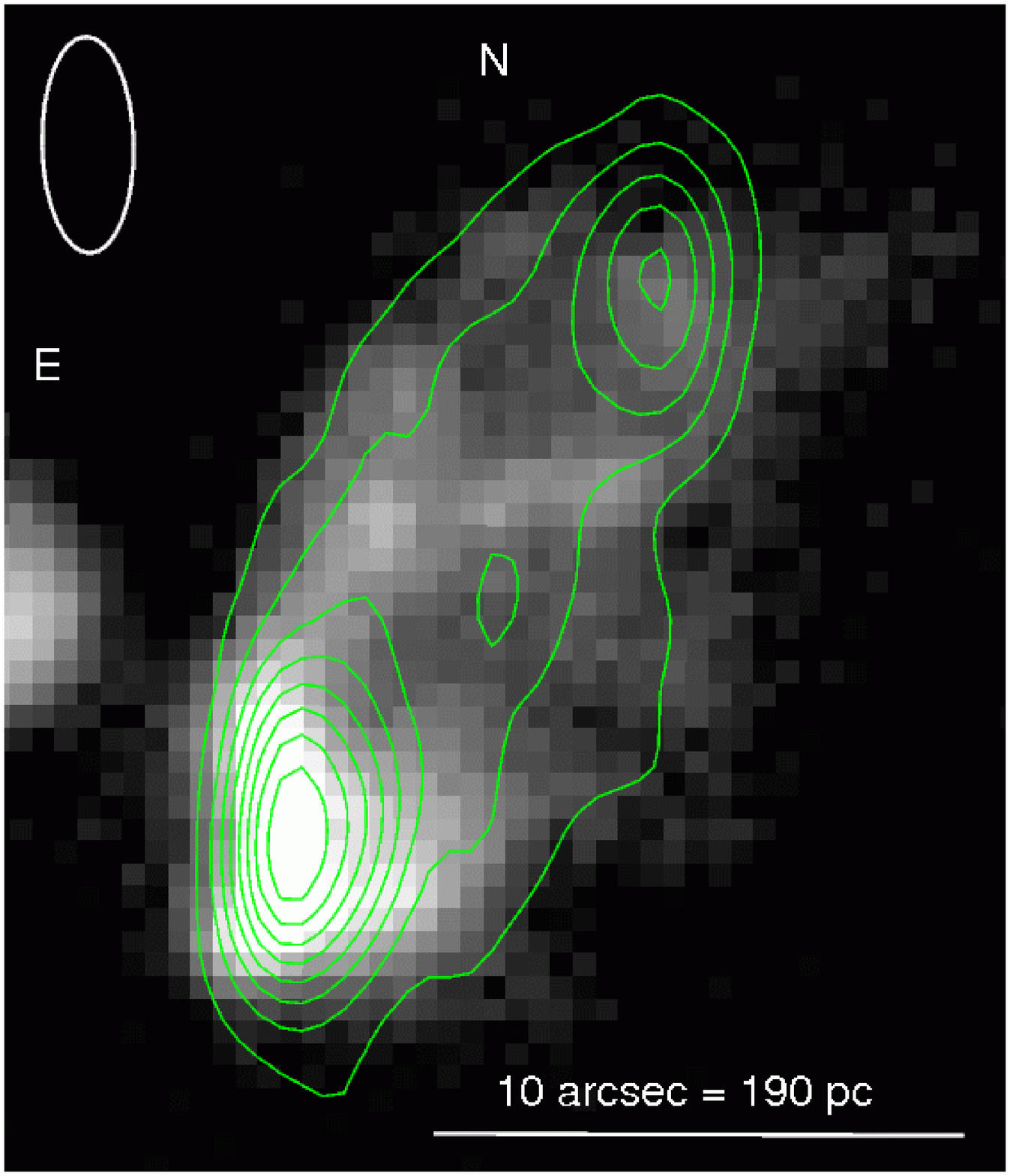}
  \caption{Left panel: {\it Chandra}/ACIS X-ray colour map, 
with ATCA 5.5-GHz radio flux contours superimposed. 
Red $= 0.3$--$1$ keV; green $= 1$--$2$ keV; blue $=2$--$8$ keV. 
Contour fluxes are at $5 \times 10^{-5}$, $10^{-4}$, $1.5 \times 10^{-4}$, 
$2 \times 10^{-4}$, $2.5 \times 10^{-4}$, $3 \times 10^{-4}$, $3.5 \times 10^{-4}$, 
$4 \times 10^{-4}$ Jy beam$^{-1}$.
Middle panel: {\it Chandra}/ACIS X-ray colour map, with 
ATCA 9-GHz radio flux contours superimposed.
Contour fluxes are at $5 \times 10^{-5}$, $7.5 \times 10^{-5}$, 
$10^{-4}$, $1.5 \times 10^{-4}$, $2 \times 10^{-4}$ Jy beam$^{-1}$.
Right panel: continuum-subtracted H$\alpha$ image 
from the 1.5m CTIO telescope; ATCA 5.5-GHz radio flux contours 
superimposed.
}
\end{figure}

\paragraph{Radio properties}

We observed the source simultaneously in the 3cm and 6cm bands 
with the Australia Telescope Compact Array, 
on 2009 Aug 6--7, for about 18 hrs. Details of the observations and data
analysis will be presented in a forthcoming paper 
(Soria et al.~2010, in prep.). We extracted 
radio maps at the effective frequencies of 5480 MHz and 8990 MHz 
(Fig.~1); the beam sizes are $4.16'' \times 1.76''$, and $2.58'' \times 1.07''$ 
respectively (Briggs robustness parameter of 0.5). 
Two bright lobes clearly stand out at the extremities of the large 
cocoon, at both frequencies. At 5.5 GHz, the peak flux in the Southern lobe 
is $\approx 0.45$ mJy beam$^{-1}$; in the Northern lobe, $\approx 0.26$ mJy beam$^{-1}$.
The total flux at 5.5 GHz is $\approx 2$ mJy, corresponding to almost 4 times 
the luminosity of Cas A. 
At 9 GHz, the peak flux in the Southern lobe 
is $\approx 0.21$ mJy beam$^{-1}$; in the Northern lobe, $\approx 0.11$ mJy beam$^{-1}$.
The spectral index (Fig.~1, right) is $\approx 0$ at the base 
of the jets, on either side of the X-ray core; it steepens 
to $\approx -0.7$ in the Southern lobe 
and $\approx -1$ in the Northern lobe.
The projected distance between the peak emission in the two lobes 
(that is, between the reverse shocks at the jet extremities) 
is $\approx 13.5'' \approx 250$ pc. 

\paragraph{X-ray properties}

From a 49-ks {\it Chandra} archival observation 
taken on 2003 Sep 7, we found an aligned triplet 
of point-like sources \citep{pak08}.
The X-ray core (Fig.~2,3) has a hard spectrum 
(photon index $\Gamma \approx 1.7$), with an emitted 
luminosity $L_{\rm 0.3-8} \approx 6 \times 10^{36}$
erg s$^{-1}$, consistent with a BH in the low/hard state 
(Pakull et al.~2010, in prep.).
The two hot spots have a much softer spectrum (Fig.~2,3), 
and are well fitted (Cash statistics $=10.4$ over 14 dof) 
by a 2-component thermal plasma model with
$kT_1 \approx 0.25$ keV and $kT_2 \approx 0.95$ keV, 
and negligible intrinsic absorption (Table 1). The emitted X-ray luminosities 
are $L_{\rm 0.3-8} \approx 5 \times 10^{36}$
erg s$^{-1}$ and $L_{\rm 0.3-8} \approx 11 \times 10^{36}$
erg s$^{-1}$ for the Northern and Southern hot spot, respectively 
(in agreement with the ratio of radio luminosities). 
%The hot spot spectra are not dominated 
%by synchrotron or synchrotron self-Compton emission. 
Simple or broken power-law models do not give acceptable fits 
(Cash statistics $=25.5$ over 15 dof); 
moreover, they would require an unphysically steep slope 
($\Gamma \approx 6$) combined with high intrinsic column densities 
($N_H \approx 5 \times 10^{21}$ cm$^{-2}$). We conclude that 
the hot spot spectra are not dominated by synchrotron or synchrotron 
self-Compton emission. Moreover, the 
projected distance between the X-ray hot spots 
is $\approx 15'' \approx 290$ pc, 
slightly larger than the distance between the radio lobes.
We interpret the X-ray hot spots as optically-thin 
thermal plasma emission from the shocked ISM gas between 
the radio lobes and the bow shock.
From the maximum size of the unresolved hot spots, we infer that 
the density of the X-ray-emitting gas is $\ga 1$ cm$^{-3}$.
A shocked-gas temperature $\approx 1$ keV corresponds to a bow shock 
velocity $v_{bs} \approx 900$ km s$^{-1}$.
%$v_{bs} \approx 300$--$400$ km s$^{-1}$. 
%(e.g., PKS\,2152$-$699: Ly et al.~2005)
%There is also diffuse X-ray emission over the whole cocoon: 
%perhaps from hot gas pervading
%the whole nebula, with average density $\approx 0.04$ cm$^{-3}$.
For an ISM of uniform density $\rho$, the distance from the core to the bow shock 
$l_{bs} \sim \rho^{-1/5}Q^{1/5}t^{3/5}$ and its velocity 
$v_{bs} \approx (3/5)(l_{bs}/t)$, 
where $Q$ is the jet power and $t$ the age of the source \citep{kai99}.
Neglecting projection effects, this suggests a characteristic age 
$\approx 10^5$ yrs and a jet power $\sim$ a few $\times 10^{40}$ 
erg s$^{-1}$.
The stronger X-ray and radio emission from the Southern hot spot and lobe, 
and their shorter distance to the core, are consistent with a higher density 
of the ISM on that side of the core.
There is also faint X-ray emission projected over the whole surface of the cocoon, 
with even softer colours than in the lobes (although the number of 
detected counts is too low for spectral fitting); this diffuse emission 
is consistent with shock velocities $\approx 300$ km s$^{-1}$. 
%It is possible that the diffuse Higher-velocity shocks in lower-density material, 
%optical shocks come from slower shocks in embedded higher-density clouds.

\begin{figure}
  \includegraphics[height=.4\textheight,angle=270]{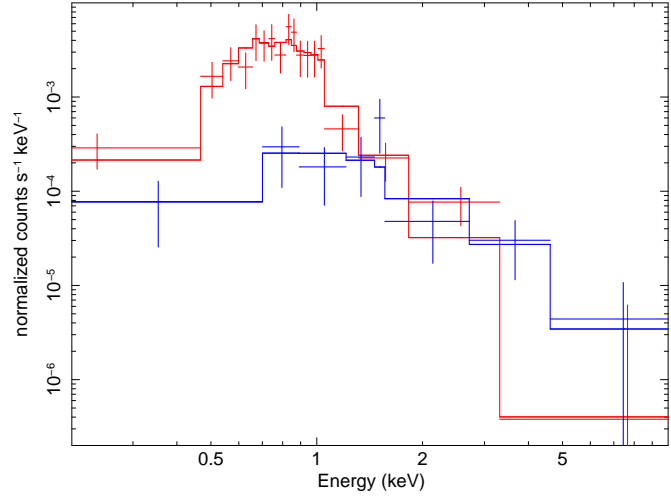}
  \caption{Red datapoints and best-fitting model: combined X-ray spectrum 
of the hot spots, fitted with a two-temperature thermal plasma model  
(best-fitting parameters listed in Table 1). Blue datapoints and best-fitting 
model: X-ray spectrum of the core, fitted with a power-law of photon index 
$\Gamma = 1.7$.}
\end{figure}

\begin{table}
\begin{tabular}{lr}
\hline
  \tablehead{1}{l}{b}{Parameter}
  & \tablehead{1}{r}{b}{Value}\\
\hline
$N_{H,\rm Gal}$ & \ \ \ \ $1.2 \times 10^{20}$ cm$^{-2}$ \ \ (fixed)\\[3pt]
$N_H$ & \ \ \ \ $< 2.9 \times 10^{20}$ cm$^{-2}$ \\[3pt]
$kT_1$ & \ \ \ \ $0.26^{+0.06}_{-0.08}$ keV\\[3pt]
$N_1$ & \ \ \ \ $\left(2.4^{+0.8}_{-0.7}\right) \times 10^{-6}$\\[3pt]
$kT_2$ & \ \ \ \ $0.96^{+0.32}_{-0.20}$ keV\\[3pt]
$N_2$ & \ \ \ \ $\left(2.0^{+0.8}_{-0.7}\right) \times 10^{-6}$\\[3pt]
\hline
$f_{0.3-8}$ & \ \ \ \ $8.8 \times 10^{-15}$ erg cm$^{-2}$ s$^{-1}$\\[3pt]
$L_{0.3-8}$ & \ \ \ \ $1.7 \times 10^{37}$ erg s$^{-1}$\\[3pt]
\hline
C-statistic &  \ \ \ \  $10.37$ using 18 PHA bins and 14 degrees of freedom.\\[3pt]
\end{tabular}
\caption{Best-fitting spectral parameters for the (combined) 
hot spot X-ray emission. The {\footnotesize {XSPEC}} model 
is {\it wabs$_{\rm Gal}$*wabs*(ray+ray)}. Errors are 90\% confidence level 
for 1 interesting parameter ($\Delta \chi^2 = 2.7$).}
\label{tab:a}
\end{table}

\paragraph{Optical properties}

Optical images from the 1.5m CTIO telescope show 
an H$\alpha$ emission nebula with similar size 
to the radio nebula (Fig.~2, right panel). From integral field 
spectra taken at the 2.3m ANU telescope, we find that the H$\alpha$ 
line is broad (full width at zero intensity $\approx 250$ km s$^{-1}$).
Moreover, using VLT data, we have recently found (Pakull et al.~2010,
in prep.) diffuse emission in the He{\footnotesize{II}} $\lambda 4686$
recombination line, with a flux $\approx 10$\% of the 
H$\beta$ flux. It is the largest known He{\footnotesize{III}} region
in the local universe. 
%Strong [OI] 6300, 6363 (at left) and strong [S II] 6717,31
Considering the relative weakness
of the central X-ray source, the most likely explanation is
shock ionization with a large shock velocity ($\approx 300$ km
s$^{-1}$). Using radiative shock model computations,
we estimate a mechanical luminosity $\approx 6 \times 10^{40}$
erg s$^{-1}$. The X-ray core has a point-like blue counterpart,
with $B \sim 23$ mag ($M_B \sim -5$ mag), consistent 
with an early-type donor star or a bright accretion disk.
The optical line emission is also stronger on the Southern side of the nebula, 
in agreement with the radio and X-ray morphology.
The shock velocity inferred from the optical spectra 
is much lower than that inferred from the X-ray temperatures. 
By analogy with well-studied supernova remnants, for example N49 in the LMC 
\citep{van92,par03}, we suggest that we have a clumpy medium, 
with the X-ray emission coming from higher-velocity shocks 
in lower-density material, and the optical lines coming from slower 
shocks in higher-density embedded clouds.

%%%%%%%%%%%%%%%%%%%%%%%%%%%%%%%%%%%%%%%%%%%%
%% Sample figure:
%%
%% The option [height=...] scales the picture to the given height,
%% without it it would be printed at its nominal size
%%%%%%%%%%%%%%%%%%%%%%%%%%%%%%%%%%%%%%%%%%%%

%%%%%%%%%%%%%%%%%%%%%%%%%%%%%%%%%%%%%%%%%%%%
%% SAMPLE TABLE
%%
%% Shows the use of \tablehead and \tablenote
%% macros
%%%%%%%%%%%%%%%%%%%%%%%%%%%%%%%%%%%%%%%%%%%%

%%%%%%%%%%%%%%%%%%%%%%%%%%%%%%%%%%%%%%%%%%%%%%%%
%% BACKMATTER
%%%%%%%%%%%%%%%%%%%%%%%%%%%%%%%%%%%%%%%%%%%%%%%%

\begin{theacknowledgments}
We thank Tasso Tzioumis for his assistance when we prepared the ATCA observations, 
and Mike Dopita for comments and for his taking of an optical spectrum for us, 
from the ANU 2.3m telescope.
RS acknowledges hospitality at Tsing Hua University (Taiwan), at Mount Stromlo Observatory, 
and at the University of Sydney, during the completion of this work.
%  Infandum, regina, iubes renovare dolorem, Troianas ut opes et
%  lamentabile regnum cruerint Danai; quaeque ipse miserrima vidi, et
%  quorum pars magna fui. Quis talia fando Myrmidonum Dolopumve aut duri
%  miles Ulixi temperet a lacrimis?
\end{theacknowledgments}



\bibliographystyle{aipproc}   % if natbib is available

%\bibliography{sample}

\begin{thebibliography}{99}

%Karachentsev, I. D.; A&A 404, 93-111 (2003)
\bibitem{sch74}
P.~A.~G. Scheuer,  \emph{MNRAS} \textbf{166}, 513--528 (1974).

\bibitem{bla74}
R.~D. Blandford, and M.~J. Rees, \emph{MNRAS} \textbf{169}, 395--415 (1974).

\bibitem{raw91}
S.~Rawlings, and R.~Saunders, \emph{Nature} \textbf{349}, 138--140 (1991).

\bibitem{kai99}
C.~R. Kaiser, and P. Alexander, \emph{MNRAS} \textbf{305}, 707--723 (1999).

%\bibitem{mei89}
%K. Meisenheimer, H.-J. Roser, P.~R. Hiltner, et al., \emph{A\&A} \textbf{219},  
%63--86 (1989).

%\bibitem{har04}
%M.~J. Hardcastle, D.~E. Harris, D.~M. Worrall, and M. Birkinshaw, \emph{ApJ} 
%\textbf{612}, 729--748 (2004).

\bibitem{kra07}
R.~P. Kraft, P.~E.~J. Nulsen, M. Birkinshaw, et al., \emph{ApJ} 
\textbf{665}, 1129--1137 (2007).

\bibitem{hei02}
S. Heinz, \emph{A\&A} \textbf{338}, L40--L43 (2002)

%\bibitem{hei06}
%S. Heinz, M.~A. Aloy, R.~P. Fender, and D.~M. Russell, 
%in \emph{Proc.~VI Microquasar Workshop: Microquasars and Beyond}, 
%Como, Italy, 2006, p.57.

\bibitem{beg80}
M.~C. Begelman, S.~P. Hatchett, C.~F. McKee, C.~L. Sarazin, and J. Arons, 
\emph{ApJ} \textbf{238},  722--730 (1980).

\bibitem{gal05}
E.~Gallo, R.~P. Fender, C.~R. Kaiser, et al., \emph{Nature} \textbf{436},
819--821 (2005).

\bibitem{mar05}
J. Mart\'i, et al.,
%D. P\'erez-Ram\'irez, J.~L. Garrido, P. Luque-Escamilla, 
%and J.~M. Paredes, 
\emph{A\&A} \textbf{439}, 279--285 (2005).

\bibitem{kai04}
C.~R. Kaiser, K.~F. Gunn, C. Brocksopp, and J.~L. Sokoloski,
\emph{ApJ} \textbf{612}, 332--341 (2004).

\bibitem{pak06}
M.~W. Pakull, F. Gris\'e, and C. Motch, 2006, 
in \emph{Populations of High Energy Sources in Galaxies}, 
proc. of the IAU Symposium 230, 
%eds E.~J.~A. Meurs and G. Fabbiano
Cambridge University Press, Cambridge, 2006, pp. 293--297.

\bibitem{pak08}
M.~W. Pakull, and F. Gris\'e,
in \emph{A Population Explosion: the Nature \& Evolution 
of X-ray Binaries in Diverse Environments}, 
AIP Conf.~Proc.~1010, AIP, 
New York, 2008, pp. 303--307.

\bibitem{gri08}
F. Gris\'e, M.~W. Pakull, R. Soria, et al., \emph{A\&A} \textbf{486}, 
151--163 (2008).

\bibitem{mil05}
N.~A. Miller,  R.~F. Mushotzky, and S.~G. Neff, 
\emph{ApJ} \textbf{623}, L109--L112 (2005).

\bibitem{sor06}
R. Soria, et al.,
%R.~P. Fender, D.~C. Hannikainen, A.~M. Read, and I.~R. Stevens, 
\emph{MNRAS} \textbf{368}, 1527--1539 (2006).

\bibitem{bla97}
W.~P. Blair, and K.~S. Long, \emph{ApJS} \textbf{108}, 261--277 (1997).

\bibitem{pan02}
T.~G. Pannuti, N. Duric, C.~K. Lacey, et al.,
\emph{ApJ} \textbf{565}, 966--981 (2002).

\bibitem{van92}
O.~Vancura, W.~P.~Blair, K.~S. Long, and J.~C. Raymond, 
\emph{ApJ} \textbf{394}, 158--173 (1992).

\bibitem{par03}
S. Park, D.~N. Burrows, and G.~P. Garmire, 
\emph{ApJ} \textbf{586}, 210--223 (2003). 



\end{thebibliography}

%  \typeout{******************************************}
%  \typeout{** Please run "bibtex \jobname" to optain}
%  \typeout{** the bibliography and then re-run LaTeX}
%  \typeout{** twice to fix the references!}
%  \typeout{******************************************}

%\end{document}

%%%%%%%%%%%%%%%%%%%%%%%%%%%%%%%%%%%%%%%%%%%
%% The following lines show an example how to produce a bibliography
%% without the help of the BibTeX program. This could be used instead
%% of the above.
%%%%%%%%%%%%%%%%%%%%%%%%%%%%%%%%%%%%%%%%%%%

\end{document}